\documentclass[runningheads]{llncs}

\usepackage{eccv}

\usepackage{eccvabbrv}

\usepackage{graphicx}
\usepackage{booktabs}

\usepackage[accsupp]{axessibility}  

\usepackage{hyperref}

\usepackage{orcidlink}
\usepackage{multirow}

\begin{document}

\title{Image Demoiréing in RAW and sRGB Domains} 
 
\titlerunning{Image Demoiréing in RAW and sRGB Domains}

\author{Shuning Xu\inst{1} \and
Binbin Song\inst{1} \and
Xiangyu Chen\inst{1,2} \and
Xina Liu\inst{1,2} \and
Jiantao Zhou\inst{1}\thanks{Corresponding Author.}}

\authorrunning{S.~Xu et al.}

\institute{State Key Laboratory of Internet of Things for Smart City, \\
Department of Computer and Information Science, University of Macau 
\and
Shenzhen Institutes of Advanced Technology, Chinese Academy of Sciences \\
\email{\{yc07425, yb97426, jtzhou\}@um.edu.mo}, \email{chxy95@gmail.com}, \email{xn.liu95@siat.ac.cn}}

\maketitle





\begin{abstract}
Moiré patterns frequently appear when capturing screens with smartphones or cameras, potentially compromising image quality. Previous studies suggest that moiré pattern elimination in the RAW domain offers greater effectiveness compared to demoiréing in the sRGB domain. Nevertheless, relying solely on RAW data for image demoiréing is insufficient in mitigating the color cast due to the absence of essential information required for the color correction by the image signal processor (ISP). In this paper, we propose to jointly utilize both RAW and sRGB data for image demoiréing (RRID), which are readily accessible in modern smartphones and DSLR cameras. We develop Skip-Connection-based Demoiréing Module (SCDM) with Gated Feedback Module (GFM) and Frequency Selection Module (FSM) embedded in skip-connections for the efficient and effective demoiréing of RAW and sRGB features, respectively. Subsequently, we design a RGB Guided ISP (RGISP) to learn a device-dependent ISP, assisting the process of color recovery. Extensive experiments demonstrate that our RRID outperforms state-of-the-art approaches, in terms of the performance in moiré pattern removal and color cast correction by 0.62dB in PSNR and 0.003 in SSIM. Code is available at \url{https://github.com/rebeccaeexu/RRID}.

\end{abstract}

\section{Introduction}
\label{sec:intro}

\begin{figure}[!ht]
\centering
\includegraphics[width=0.6\linewidth]{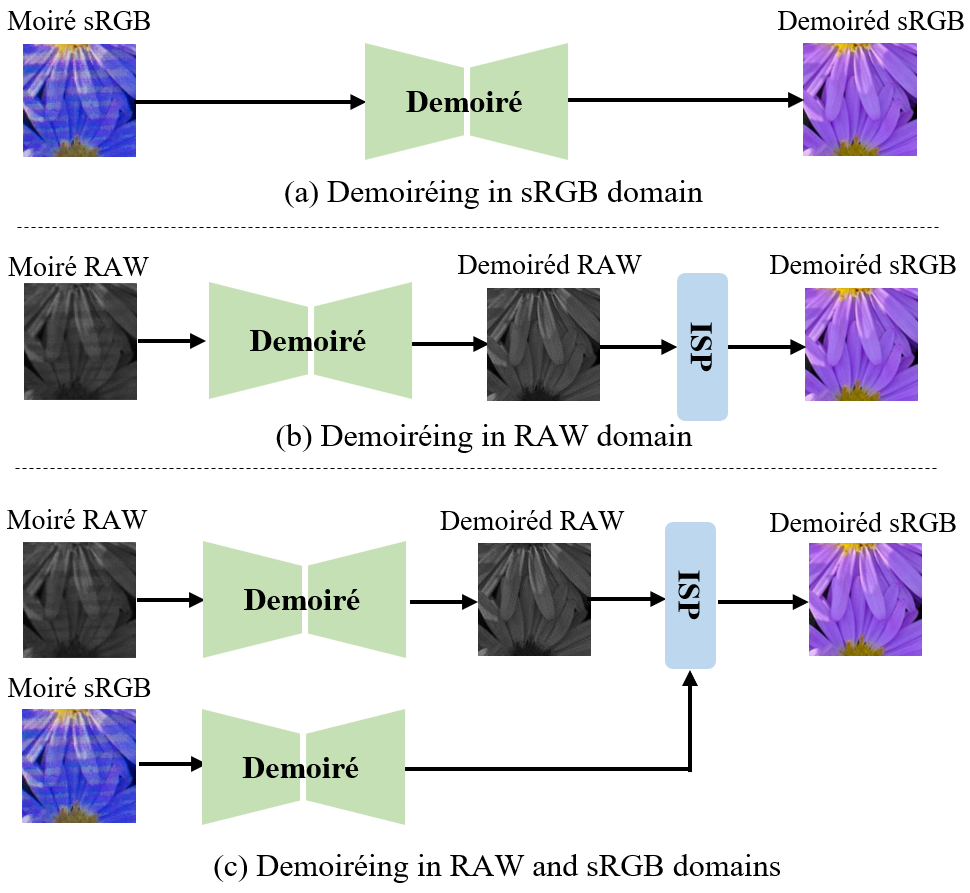}
\caption{Comparison of image demoiréing in different domains. (a) Image demoiréing in sRGB domain. (b) Image demoiréing in RAW domain. (c) Our proposed RRID, which performs image demoiréing in both RAW domain and sRGB domain}
\label{fig:introduction}
\end{figure}

With the increasing prevalence of smartphone cameras, it is common to record on-screen content using smartphone cameras for convenience.
However, this type of photography often results in the presence of moiré patterns in the captured images. The appearance of moiré patterns is caused by the spatial frequency aliasing between the camera's color filter array (CFA) and the screen's LCD subpixel~\cite{wang2023coarse}, resulting in an unsatisfactory visual experience. 

Moiré patterns present challenges to be eliminated due to their different scales, indistinct shapes, diverse colors, and varying frequencies. Nowadays, many learning-based image demoiréing approaches~\cite{sun2018moire, cheng2019multi, he2019mop, he2020fhde, liu2020wavelet, zheng2020image, liu2020mmdm, wang2021image, yu2022towards, niu2023progressive, wang2023coarse} have been proposed. Among the various methods employed, a commonly used approach is utilizing a multi-scale architecture to remove various sizes of moiré patterns~\cite{sun2018moire, cheng2019multi, he2019mop, yu2022towards, niu2023progressive}.
Besides, some works advocate the leverage of frequency components and delicately design frequency-based demoiréing networks~\cite{he2020fhde, liu2020wavelet, zheng2020image, wang2023coarse}. 

The aforementioned methods are developed to demoiré in the sRGB domain, as depicted in Fig.~\ref{fig:introduction}(a). While this scheme seems to be the most direct approach, its effectiveness is limited in removing moiré patterns from complex scenes. The reason for this limitation probably lies in the fact that the nonlinear operations in ISP such as demosaicing deteriorate the moiré patterns originally in the RAW domain. Many schemes consequently suggest that eliminating moiré patterns in the RAW domain is more effective than performing demoiréing in the sRGB domain~\cite{yue2022recaptured, yue2023recaptured}. Due to the easy accessibility of RAW data in modern smartphones and DSLR cameras, image demoiréing in RAW domain can be both feasible and advantageous.
RDNet~\cite{yue2022recaptured} is the pioneering study that investigates image demoiréing in the RAW domain. They explore the demoiréing of RAW images using an encoder-decoder and class-specific learning approach, followed by the utilization of a pre-trained ISP for the RAW-to-sRGB conversion, as depicted in Fig.~\ref{fig:introduction}(b). Nevertheless, relying exclusively on RAW data can lead to color cast due to the uncertainty during RAW-to-sRGB conversion. 

In order to correct the color cast and make the best use of RAW and sRGB information for the moiré pattern removal, we propose to jointly utilize both RAW and sRGB data for image demoiréing (RRID), as demonstrated in Fig.~\ref{fig:introduction}(c). We advocate that introducing paired RAW-sRGB data is advantageous to the moiré pattern removal due to the following reasons: (1) RAW pixels provide more information compared to sRGB pixels since they are typically 12 or 14 bits. (2) Moiré patterns in the RAW domain are less apparent because they are not further affected by nonlinear operations in the ISP. (3) The paired RAW and sRGB data allows the model to learn a device-dependent ISP to aid the process of color recovery. 
In RRID, we adopt a multi-scale architecture in the Skip-Connection-based Demoiréing Module (SCDM), to effectively eliminate moiré patterns of different sizes. In order to effectively perform demoiréing on RAW and sRGB features, the Gated Feedback Module (GFM) and Frequency Selective Module (FSM) are integrated into skip-connections within SCDM, as opposed to brutally injecting them into the multi-scale layers. Specifically, for RAW features demoiréing, we introduce the GFM into the skip connection, enabling adaptive differentiation between texture details and moiré patterns through feature gating. Moreover, we design the FSM for the moiré sRGB features, leveraging a learnable band stop filter to mitigate moiré patterns in the frequency domain. Subsequent to the pre-demoiréing operations, we develop the RGB Guided Image Signal Processor (RGISP) to integrate color information from the demoiréd sRGB features with the demoiréd RAW features and learn a device-dependent ISP, facilitating the color recovery process. Ultimately, for the image reconstruction, we utilize multiple Residual Swin Transformer Blocks (RSTBs)~\cite{liang2021swinir} and convolutions to accomplish global tone mapping and detail refinement.

In summary, our contributions are listed as follows:
\begin{itemize}
\item[$\bullet$] We propose to exploit image demoiréing in both RAW and sRGB domains. The incorporation of RAW data enhances the moiré pattern removal process, while the inclusion of sRGB data facilitates the RAW-to-sRGB conversion and aids in color restoration of images.
\item[$\bullet$] We develop SCDM with specifically designed modules, GFM and FSM, embeded in the skip-connections to perform efficient and effective demoiréing on the RAW and sRGB features.
\item[$\bullet$] RGISP is designed for the RAW-to-sRGB conversion, incorporating color information from the demoiréd sRGB features with demoiréd RAW features and learn a device-dependent ISP, aiding the color deviation correction.
\item[$\bullet$] Our RRID surpasses state-of-the-arts methods in both qualitative and quantitative evaluations.
\end{itemize}

\section{Related Works}
\label{sec:related}

\subsection{Image and Video Demoiréing}
The purpose of image demoiréing is to restore the original clean image by removing moiré patterns and correcting color deviations from an contaminated image. 
Many learning-based image demoiréing~\cite{cheng2019multi, he2020fhde, zheng2020image, liu2020mmdm, wang2021image, niu2023progressive, wang2023coarse} and video demoiréing methods~\cite{dai2022video, oh2023fpanet, xu2024direction, yue2023recaptured} have been introduced to mitigate the moiré patterns.
DMCNN~\cite{sun2018moire} constructs the first real-world dataset for image demoiréing and proposes a multi-resolution neural network to handle moiré patterns across various scales.
MopNet~\cite{he2019mop} devises a multi-scale aggregated network that leverages the edge guidance and pattern attributes for moiré pattern removal.
WDNet~\cite{liu2020wavelet} employs wavelet transformation to decompose images with moiré patterns into separate frequency bands and establishes a dual-branch network to restore both close-range and far-range information.
FHDMi~\cite{he2020fhde} formulates a two-stage approach for simultaneous removal of substantial moiré patterns while preserving image details.
ESDNet~\cite{yu2022towards} explores a lightweight model aimed at ultra-high-definition image demoiréing.
VDMoiré~\cite{dai2022video} collects the first video demoiréing dataset and presents a baseline video demoiréing model with a relation-based temporal consistency loss. 
The aforementioned methods all operate in the sRGB domain. Since it has been observed that moiré patterns are less prominent in the RAW domain, some schemes advocate performing image or video demoiréing in the RAW domain.
RDNet~\cite{yue2022recaptured} introduces the first image demoiréing dataset with RAW data and conducts the demoiréing process in the RAW domain.
RawVDemoiré~\cite{yue2023recaptured} proposes a temporal alignment method for RAW video demoiréing.
To remove moiré patterns in the RAW domain, the model have to simultaneously possess the capabilities of demoiréing and and RAW-to-sRGB conversion. However, previous methods encounter challenges in color recovery due to the uncertainty of RAW-to-sRGB conversion.
Consequently, we propose an image demoiréing network that utilizes paired RAW-sRGB data, facilitating the color recovery process in image demoiréing.

\subsection{Learning-Based RAW Image Processing}
RAW pixels offer the potential for leveraging additional information due to their broader bit depth, inherently containing a great wealth of data. Early research in RAW image processing primarily centers on integrating demosaicing and denoising techniques~\cite{khashabi2014joint, gharbi2016deep, kokkinos2018deep, liu2020joint}.
RTF~\cite{khashabi2014joint} devices a machine learning approach to the demosaicing and denoising on RAW inputs.
SGNet~\cite{liu2020joint} introduces a self-guidance network that leverages the inherent density-map guidance and green-channel guidance for demosaicing and denoising on the RAW images.
Recently, some low-level vision tasks endorse the utilization of RAW data, including low-light enhancement~\cite{huang2022towards, dong2022abandoning, jin2023dnf}, super-resolution~\cite{xing2021end, yue2022real, luo2024and}, and reflection removal~\cite{lei2021robust, song2023real}.
The majority of the aforementioned RAW image processing methods only take RAW images as input and generate restored sRGB images as output, indicating that the entire model is responsible for both the restoration and ISP tasks.
CR3Net~\cite{song2023real} presents a cascaded network that harnesses both the RGB data and their corresponding RAW versions for image reflection removal. Within CR3Net, a module is formulated to emulate point-wise mappings within the ISP, converting the features from RAW to sRGB domain. 
To further mitigate the color casting problem, our proposed method jointly utilize RAW and sRGB data to learn a device-dependent ISP for image demoiréing.

\section{Methodology}
\label{sec:method}
Originally caused by frequency interference between the camera’s CFA and the screen’s LCD subpixel, moiré patterns are further deteriorated by the nonlinear processes in ISP. 
Therefore, introducing RAW data for image demoiréing may offer greater effectiveness. However, relying solely on RAW data for image demoiréing is insufficient in mitigating color cast due to the uncertainty during the RAW-to-sRGB conversion. Consequently, in this work, we propose to utilize paired RAW and sRGB data, which are readily accessible in modern smartphones and DSLR cameras (\eg, iPhone 15 Pro, HUAWEI P60 Pro). The RAW-sRGB pairs essentially allows the model to learn a device-dependent ISP, aiding the color recovery process. The potential applications in real-world scenarios can be described as follows. When demoiréing function is enabled, the in-device ISP generated sRGB image will not be directly outputted. Instead, together with its corresponding RAW version, will be processed by our RRID to generate a new sRGB image with much suppressed moiré patterns. Afterwards, the RAW image could be discarded, depending on settings of the camera. 
\begin{figure*}[!ht]
\centering
\includegraphics[width=1\linewidth]{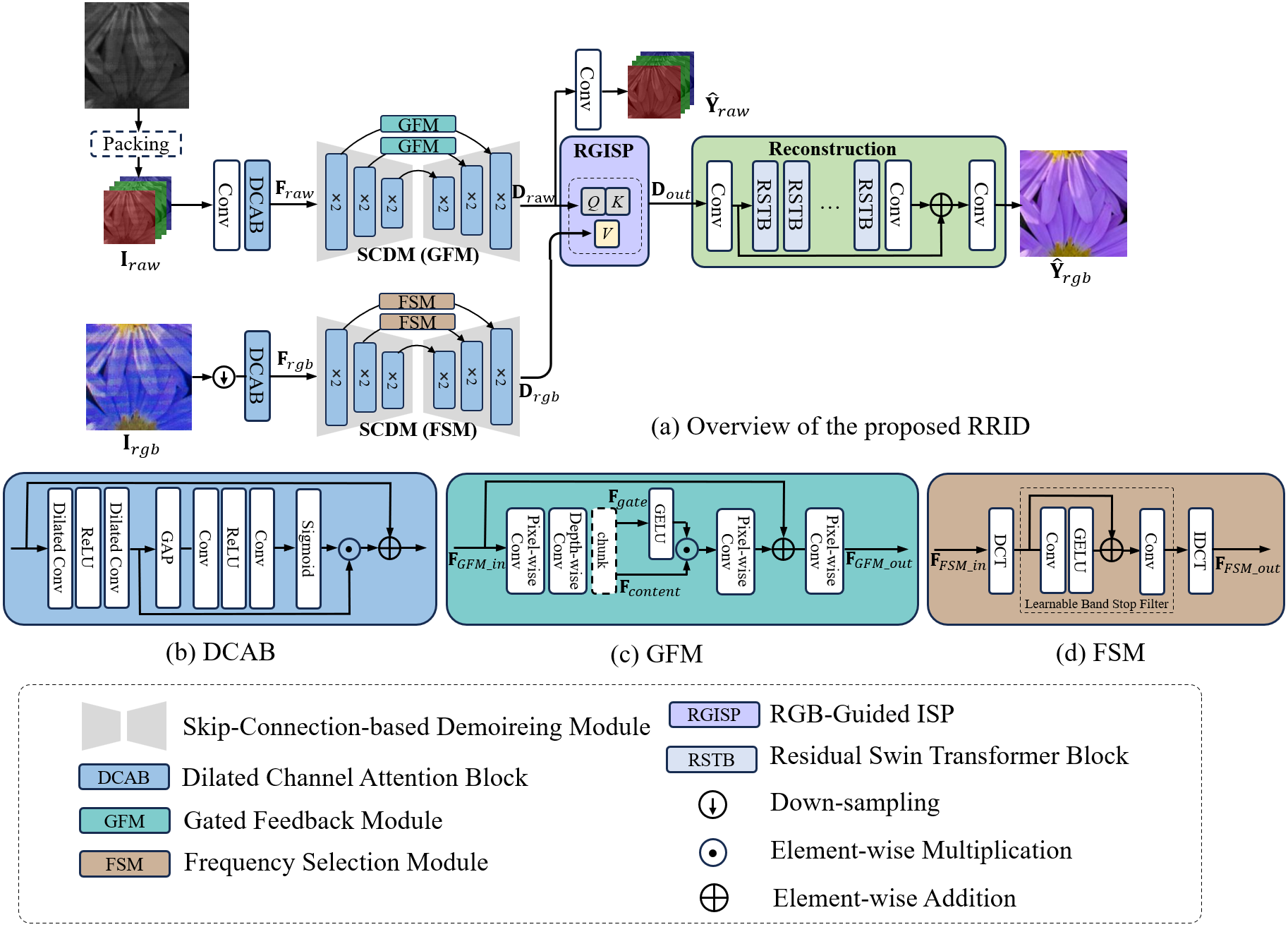}
\caption{The overview and detailed structures of our proposed RRID.}
\label{fig:architecture}
\end{figure*}


More specifically, Fig.~\ref{fig:architecture} presents an overview of our proposed RRID for image demoiréing with the paired RAW and sRGB data. As for the RGB branch,  the input is an sRGB image $\mathbf{I}_{rgb}\in \mathbb{R}^{H\times W \times 3}$, where $H \times W$ represents the spatial resolution. Given an input RAW image, it is first packed into the 4-channel RGGB format $\mathbf{I}_{raw}\in \mathbb{R}^{\frac{H}{2}\times \frac{W}{2} \times 4}$, which serves as the input for the RAW branch. 
RRID is designed to generate the final output sRGB image ($\mathbf{\hat{Y}}_{rgb}$).

Concretely, shallow features $\mathbf{F}_{raw}$ and $\mathbf{F}_{rgb}$ are respectively extracted from the input data $\mathbf{I}_{raw}$ and $\mathbf{I}_{rgb}$ using a convolutional layer and the Dilated Channel Attention Block (DCAB). It is worth noting that a down-sampling layer is applied to the sRGB data to ensure the uniformity of feature shapes.
Then, we apply SCDM to acquire pre-demoiréd features $\mathbf{D}_{raw}$ and $\mathbf{D}_{rgb}$ in RAW branch and RGB branch, respectively.
GFM is introduced in the skip-connection of SCDM for RAW features, enabling adaptive differentiation between texture details and moiré patterns through feature gating. Similarly, FSM is utilized to to mitigate moiré patterns in the frequency domain. More details will be given in Sec.~\ref{sec:method_scdm}.
Subsequently, RGISP is designed to realize a RAW-to-sRGB conversion, generating $\mathbf{D}_{out}$. RGISP incorporates color information from coarsely demoiréd sRGB features $\mathbf{D}_{rgb}$ in a cross-attention mechanism, which will be introduced in Sec.~\ref{sec:method_rgisp}.
For the final image reconstruction, 4 RSTBs~\cite{liang2021swinir} are utilized to generate sRGB output $\mathbf{\hat{Y}}_{rgb}$ for their advantages in building long-range dependencies for global tone mapping and detail refinement.

\subsection{Design of SCDM}
\label{sec:method_scdm}
We propose SCDM, a multi-scale architecture designed to remove moiré patterns from shallow RAW features $F_{raw}$ and shallow sRGB features $F_{rgb}$, as shown in Fig.~\ref{fig:architecture}(a). Unlike the vanilla U-Net, SCDM is constructed upon DCAB, facilitating the encoding and decoding of informative features within the respective layers. The features are downsampled twice in the spatial dimension during the encoding process and then upsampled to the original resolution of the input features during decoding. To emphasize the contextual information, we enhance the learned features from the corresponding layers of the encoder with additional modules at skip connections, that is GFM for RAW and FSM for sRGB. The introduction of GFM and FSM in skip-connections is also intended to enhance model efficiency. Then, we can obtain the pre-demoiréd features $\mathbf{D}_{raw}$ and $\mathbf{D}_{rgb}$.

\textbf{DCAB.}
DCAB is the essential building block of encoder-decoders for performing RAW demoiréing and the RGB demoiréing, as shown in Fig.~\ref{fig:architecture}(b). DCAB is designed using a series of dilated convolution layers followed by ReLU activation and the channel attention mechanism. The receptive field at each block is expanded with the dilated convolution kernels, facilitating the demoiréing operations. Furthermore, by incorporating the channel attention mechanism, DCAB adaptively rescales features according to interdependencies among channels. Also, shortcuts are introduced to establish connections between the input and output of DCAB, enabling the block to learn the residual features. 

\textbf{GFM.}
Embeded in the skip-connection of SCDM for RAW features, GFM is introduced to enable adaptive differentiation between the texture and moiré patterns through feature gating. The detailed structure of the GFM is shown in Fig.~\ref{fig:architecture}(c). For efficiency, we employ point-wise and depth-wise convolutions to aggregate channel and local content information, respectively. Then, the features are chunked along the channel dimension to generate $\mathbf{F}_{gate}$ and $\mathbf{F}_{content}$. A point-wise multiplication is applied on $\mathbf{F}_{content}$ and $\mathbf{F}_{gate}$ after a $\mathrm{GELU}$ activation.
During feature gating, we expect that the original image content is adaptively selected and merged along both spatial and channel dimensions. The operations can be formulated as:
\begin{gather}
    \{\mathbf{F}_{gate}, \mathbf{F}_{content}\} = \mathrm{DConv}(\mathrm{PConv}(\mathbf{F}_{GFM\_in})), \\
    \mathbf{F}_{GFM\_out} = \mathrm{PConv}(\mathbf{F}_{content} \odot \mathrm{GELU}(\mathbf{F}_{gate}) + \mathbf{F}_{GFM\_in}),
\end{gather}
where $\mathrm{DConv}$ and $\mathrm{PConv}$ represent depth-wise convolution and point-wise convolution, respectively. Here, $\odot$ denotes the element-wise multiplication.

\textbf{FSM.}
For sRGB demoiréing, we leverage a learnable band stop filter~\cite{zheng2020image} to mitigate moiré patterns in the frequency domain.
Considering the formation of moiré patterns is caused by the spatial frequency aliasing between the camera CFA and screen's LCD subpixel, demoiréing in the frequency domain becomes a favorable approach.
However, the frequency spectrum of the moiré pattern is often mixed with that of the original contents, thereby making it challenging to disentanglement. Therefore, we propose to perform the disentanglement within small patches, where the moiré patterns tend to be more pronounced in a specific range of frequency bands, facilitating the removal of moiré patterns. Inspired by Zheng \etal~\cite{zheng2020image}, we utilize a band stop filter to amplify certain frequencies and suppress others using Block DCT. However, obtaining the frequency prior to separate moiré patterns from normal image textures is challenging and time-consuming. To address this issue, a set of convolution layers are employed as the learnable band stop filter $\mathrm{B}(\cdot)$ to attenuate the specific frequencies of moiré patterns while preserving the original image contents. The process of the FSM can be represented as:
\begin{equation}
    \mathbf{F}_{FSM\_out} = \mathrm{IDCT}( \mathrm{B}( \mathrm{DCT}(\mathbf{F}_{FSM\_in}))),
   \label{equ:fsm}
\end{equation}
where $\mathrm{DCT}$ and $ \mathrm{IDCT}$ denote the process of Block DCT and the corresponding inverse process. The block size is set to $8\times8$ for $\mathrm{DCT}$ and $\mathrm{IDCT}$.

\subsection{Design of RGISP}
\label{sec:method_rgisp}

\begin{figure}[!ht]
\centering
\includegraphics[width=0.8\linewidth]{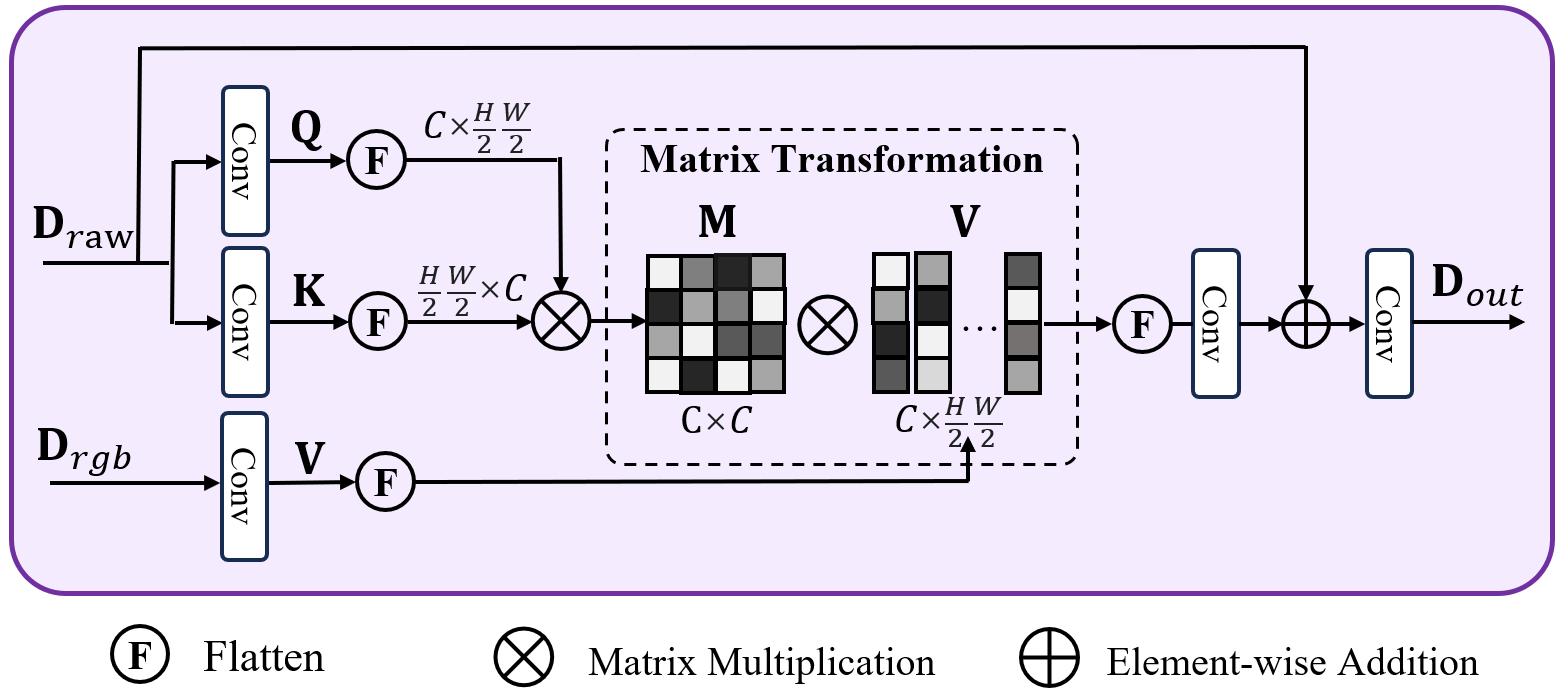}
\caption{The structure of RGISP.}
\label{fig:RGISP}
\end{figure}

RGISP is proposed to transform pre-demoiréd RAW features $\mathbf{D}_{raw}$ into sRGB domain $\mathbf{D}_{out}$. With the assistance of pre-demoiréd sRGB features $\mathbf{D}_{rgb}$, a device-dependent ISP can be learned in RGISP.
Matrix transformation is commonly employed in the traditional ISP pipelines~\cite{nakamura2017image}. It enables the enhancement or conversion of image colors to another color space through a channel-wise matrix transformation, which is facilitated by globally shared settings such as environmental illumination and color space specifications~\cite{jin2023dnf}. Building upon this principle, we introduce matrix transformation as a means to perform color transformation, as illustrated in Fig.~\ref{fig:RGISP}. The design of this matrix transformation is motivated by recent advancements in transposed cross-attention.
Given $\mathbf{D}_{raw} \in \mathbb{R}^{C\times \frac{H}{2} \times \frac{W}{2}}$, the vectors of query $\mathbf{Q}\in \mathbb{R}^{C\times \frac{H}{2}\frac{W}{2}}$, key $\mathbf{K}\in \mathbb{R}^{\frac{H}{2}\frac{W}{2}\times C}$ are generated through the projection with a 1 $\times$ 1 convolutional layer $\mathrm{Conv}$ and a flatten operation $\mathrm{Flatten}$, where $C$ is the intermediate channel number. Similarily, value $\mathbf{V}\in \mathbb{R}^{C\times \frac{H}{2}\frac{W}{2}}$ can be produced by operations on $\mathbf{D}_{rgb} \in \mathbb{R}^{C\times \frac{H}{2} \times \frac{W}{2}}$ as:
\begin{gather}
    \{\mathbf{Q},\mathbf{K}\} = \mathrm{Flattern}(\mathrm{Conv}(\mathbf{D}_{raw})), \\
    \mathbf{V} = \mathrm{Flatten}( \mathrm{Conv}(\mathbf{D}_{rgb})).
\end{gather}
Then, the transformation matrix $\mathbf{M}\in \mathbb{R}^{C\times C}$ is obtained by matrix multiplication. The procedure can be formulated as:
\begin{equation}
    \mathbf{M} = \mathrm{Softmax}(\mathbf{Q} \cdot \mathbf{K}^T/ \lambda),
   \label{equ:rgisp_3}
\end{equation}
where a scaling coefficient $\lambda$ is applied for numerical stability. Subsequently, the vector $\mathbf{V}$ is transformed by the matrix $\mathbf{M}$, performing color space conversion in feature-level. The output feature after color transformation can be obtained by $\mathbf{D}_{out} = \mathbf{M} \cdot \mathbf{V}$. As a complement to the global matrix transformation, we use a depth-wise convolution and a point-wise convolution to refine the local details, generating the output in sRGB domain $\mathbf{D}_{out}$.

\subsection{Loss Function}
\label{sec:method_loss}
To effectively remove the moiré patterns and adjust the color deviations, we introduce supervision in both RAW and sRGB domains. The overall training objective can be expressed as:
\begin{equation}
\small
\label{eq:loss}
\mathcal{L} = \alpha \|\hat{\mathbf{Y}}_{raw} - \mathbf{Y}_{raw}\|_1 + \|\hat{\mathbf{Y}}_{rgb} - \mathbf{Y}_{rgb}\|_1,
\end{equation}
where ${\mathbf{Y}}_{raw}$ and ${\mathbf{Y}}_{rgb}$ are the ground-truth RAW and sRGB images, respectively. We empirically set $\alpha$=0.5.

\section{Experiments}
\label{sec:exp}

\subsection{Experimental Setup}
\textbf{Dataset.}
The experiments are conducted on the RAW image demoiréing dataset published in TMM22~\cite{yue2022recaptured}, which contains recaptured scenes of natural images, documents, and webpages. In TMM22, the dataset consists of 540 RAW and sRGB image pairs with ground-truth for training, and 408 pairs for testing. To facilitate the training and comparison process, patches of size 256$\times$256 and 512$\times$512 are cropped respectively for the training and testing sets. The moiré images are captured using four different smartphone cameras and three display screens of varying sizes.
In addition, we utilize an image demoiréing dataset FHIMi~\cite{he2020fhde} to verify the generalization ability of our method. FHDMi contains 9981 sRGB image pairs for training and 2019 for testing, featuring diverse and intricate moiré patterns. Notably, images in FHDMi have a higher resolution of 1920$\times$1080, compared to TMM22.

\noindent\textbf{Training Details.}
We train RRID using the AdamW optimizer with $\beta_1$ = 0.9 and $\beta_2$ = 0.999. A multistep learning rate schedule is employed, with the learning rate initialized at 2$\times10^{-4}$. The RRID model is trained for 500 epochs using a batch size of 80 on 4 NVIDIA RTX 3090 GPUs.

\subsection{Quantitative Results}
We compare our approach with several image demoiréing methods in sRGB domain, including DMCNN~\cite{sun2018moire}, WDNet~\cite{liu2020wavelet}, ESDNet~\cite{yu2022towards}, and a RAW image demoiréing method RDNet~\cite{yue2022recaptured}. We further add comparison with a low-light enhancement method DNF~\cite{jin2023dnf} with RAW input, a video demoiréing method RawVDmoiré~\cite{yue2023recaptured} with RAW input, and a de-reflection method CR3Net~\cite{song2023real} with paired RAW-sRGB data. To evaluate the performance of our proposed RRID, we adopt the following three standard metrics to assess pixel-wise accuracy and the perceptual quality: PSNR, SSIM~\cite{wang2004image}, and LPIPS~\cite{zhang2018unreasonable}. Additionally, the number of parameters, FLOPs, and inference time are introduced to measure the model complexity. To ensure a fair comparison, we modify the inputs of the sRGB-based models to align with the dataset requirements. Subsequently, all models are fine-tuned using the default settings as provided in their papers on TMM22 dataset. We select the better results from the pretrained and retrained models for comparison, which provides an advantage over the competing methods.
\begin{table}[htbp]
    \centering
    \caption{Quantitative comparison with the state-of-the-art demoiréing approaches and RAW image restoration methods on TMM22 dataset \cite{yue2022recaptured} in terms of average PSNR, SSIM, LPIPS and computational complexity. The best results are highlighted with \textbf{bold}. The second-best results are highlighted with \underline{underline}.}
    \label{table:quantitative-rawid}
    \scalebox{0.83}{
        \begin{tabular}{ccccccccc}
        \toprule
        \multirow{3}*{Index} & \multicolumn{8}{c}{Methods}\\
		\cmidrule(r){2-9}
		 ~ & DMCNN & WDNet  & ESDNet & RDNet & DNF & RawVDmoiré & CR3Net & RRID\\
         ~ & \cite{sun2018moire} & \cite{liu2020wavelet} & \cite{yu2022towards} & \cite{yue2022recaptured} & \cite{jin2023dnf} & \cite{yue2023recaptured} & \cite{song2023real} & (Ours)\\
        \midrule
        \# Input type & sRGB & sRGB & sRGB & RAW & RAW & RAW & sRGB+RAW & sRGB+RAW \\
        \midrule
        PSNR$\uparrow$ & 23.54 & 22.33 & 26.77 & 26.16 & 23.55 & \underline{27.26} & 23.75 &  \textbf{27.88} \\
        SSIM$\uparrow$ & 0.885 & 0.802 &  0.927 & 0.921 & 0.895 & \underline{0.935} & 0.922 & \textbf{0.938} \\
        LPIPS$\downarrow$ & 0.154 & 0.166 & 0.089 & 0.091 & 0.162 & \textbf{0.075} & 0.102 & \underline{0.079} \\
        \cmidrule(r){1-9}
        Params (M) & \underline{1.55} & 3.36 & 5.93 & 6.04 & \textbf{1.25} & 5.33 & 2.68 & 2.38\\
        TFLOPs & 0.102 & 0.055  & 0.141 & 0.161 & \textbf{0.013} & \underline{0.022} & 0.883 & 0.093 \\
        Inference time (s) & \textbf{0.052} & 0.284 & 0.115 & 1.094 & 0.070 & 0.182 & \underline{0.058} & 0.089 \\
        \bottomrule
        \end{tabular}
    }
\end{table}

Table.~\ref{table:quantitative-rawid} presents the quantitative and the computational complexity comparisons. RRID outperforms the second-best method, RawVDmoiré, by 0.62dB in PSNR and 0.003 in SSIM. As for LPIPS, our methods accomplishes the second-best result at 0.079, surpassing most of the comparison methods. 
Even though RRID utilizes both RAW and sRGB data as inputs, our approach can still achieve a satisfactory demoiréing effect with a inference time of merely 0.089s. Compared to previous methods, our RRID demonstrates superiority with acceptable parameters and FLOPs, further demonstrating the superiority of our approach.

\begin{table}[htbp]
\caption{Quantitative comparison on FHDMi dataset~\cite{he2020fhde}.}
\label{table:quantitative-fhdmi}
\centering
    \scalebox{0.83}{
        \begin{tabular}{lccc} 
            \toprule
            {Method}  & PSNR$\uparrow$ & SSIM$\uparrow$ \\\midrule
            DMCNN~\cite{sun2018moire} & 21.54 & 0.773 \\
            MopNet~\cite{he2019mop} & 22.76 & 0.796 \\
            MBCNN~\cite{zheng2020image} & 22.31 & 0.810 \\
            FHDe$^2$Net~\cite{he2020fhde} & 22.93 &  0.789 \\
            ESDNet~\cite{yu2022towards} & \textbf{24.50} &  \textbf{0.835} \\
            Ours & \underline{24.39} & \underline{0.830} & \\
            \bottomrule
        \end{tabular}}
\end{table}

To verify the generalization performance of our method, we conduct additional experiments on an image demoiréing dataset FHIMi~\cite{he2020fhde}. In order to adapt to the FHDMi dataset with sRGB data as input, the RAW branch and RGISP are removed. We select methods tailored for image demoiréing: DMCNN~\cite{sun2018moire}, MopNet~\cite{he2019mop}, MBCNN~\cite{zheng2020image}, FHDe$^2$Net~\cite{he2020fhde}, and ESDNet~\cite{yu2022towards} for comparison, as shown in Table \ref{table:quantitative-fhdmi}.
Although RRID is initially designed for demoiréing on the RAW-sRGB image pairs, it still achieves the second-best performance by 24.39dB in PSNR and 0.830 in SSIM.
Compared to ESDNet, an image demoiréing network designed for ultra-high-definition sRGB images with more than twice of the parameters, RRID also yields competitive results.
The qualitative results on the FHDMi dataset demonstrate that our method can not only perform well in demoiréing on RAW-sRGB pairs but also solve sRGB image demoiréing, exhibiting a certain degree of generalization.

\subsection{Qualitative Results}
\begin{figure*}[!ht]
\centering
\includegraphics[width=1\linewidth]{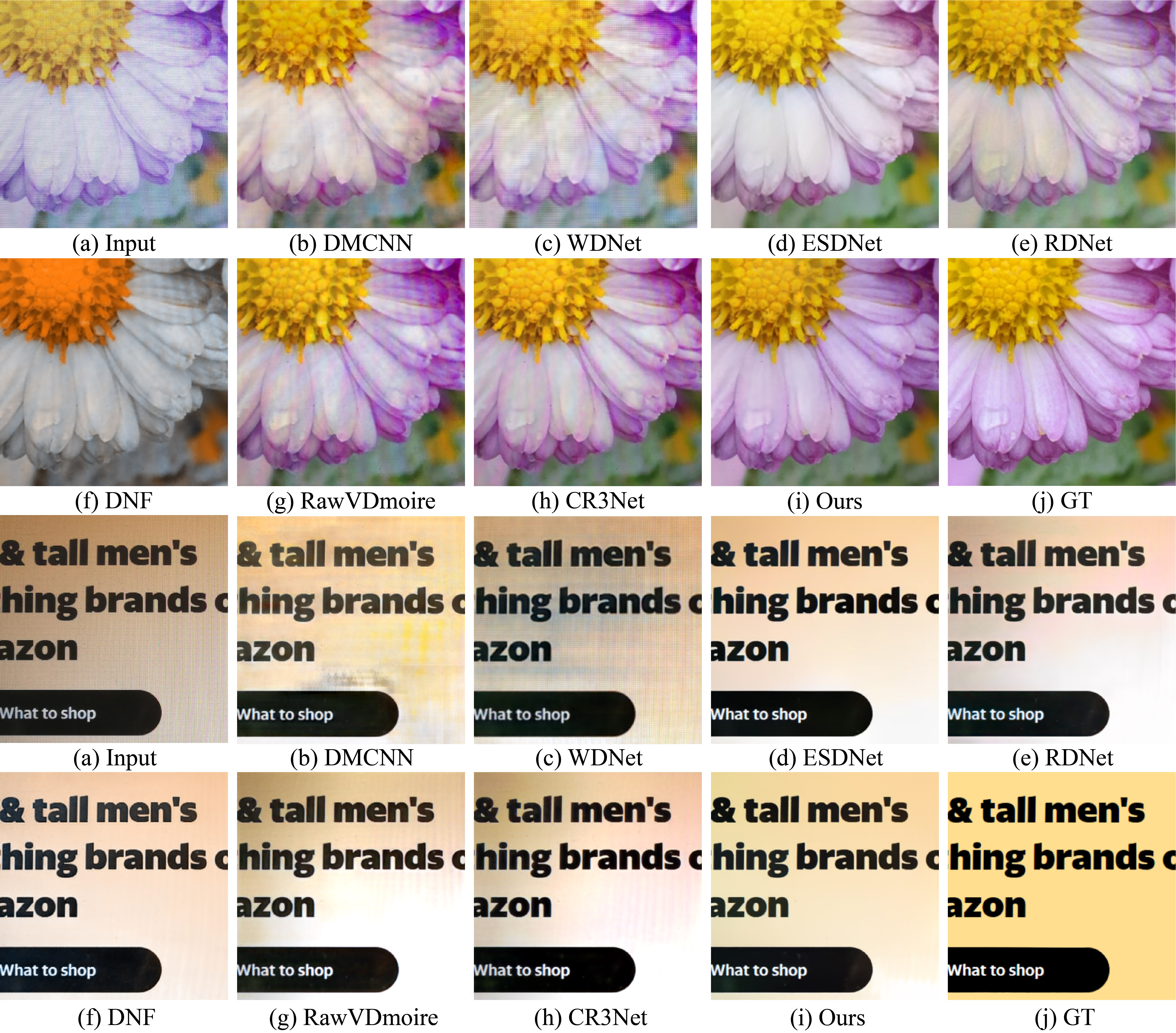}
\caption{Qualitative comparison on RAW image demoiréing TMM22 dataset~\cite{yue2022recaptured}.}
\label{fig:TMM22}
\end{figure*}

We present visual comparisons among our approach and the existing methods in Fig.~\ref{fig:TMM22}. 
The results demonstrate the advantage of RRID in effectively removing moiré artifacts and correcting color deviations. 
For the first scene demonstrating a recaptured flower with moiré patterns, previous methods tend to exhibit residual moiré patterns in their restored images. While ESDNet successfully eliminates a significant portion of these patterns, it struggles to accurately restore the original colors of the image.
Apart from natural scenes, we also present a case of webpages contaminated with moiré patterns. Through the utilization of both RAW and sRGB data, our approach eliminates moiré patterns while preserving fine image details, even in scenarios characterized by severe color deviations. This ensures a more accurate restoration of the original colors and results in enhanced visual effects.
Additional qualitative comparisons can be found in the supplementary file.

\subsection{Ablation Study}
\noindent\textbf{1) Ablation study on the model architecture and inputs.}
In Table~\ref{table:ablation_baseline}, we investigate the architecture and inputs of RRID and validate the importance of different individual components in the whole RRID. We conduct this ablation study by comparing the proposed RRID and the following variants of RRID:
\begin{itemize}
\item[$\bullet$] $B_1$. Both the RAW input and the RAW demoiréing branch are removed. Due to the lack of RAW features, the entire RGISP is also deleted. 
\item[$\bullet$] $B_2$. Both the sRGB input and sRGB demoiréing branch are removed.
\item[$\bullet$] $B_3$. The sRGB input is substituted with RAW input. The input channel number is adjusted, while the rest of the architecture is kept unchanged.
\item[$\bullet$] $B_4$. We replace RGISP with a two convolutional layers after a concatenation of the RAW and sRGB features.
\item[$\bullet$] $B_5$. Reconstruction module is replaced with two convolutional layers. 
\item[$\bullet$] $B_6$. The full model of RRID with both RAW and sRGB images as input. 
\end{itemize}

\begin{table}[htbp]
\caption{Comparison of baseline models for the evaluation of architecture and inputs.}
\label{table:ablation_baseline}
\centering
    \scalebox{0.83}{
        \begin{tabular}{lccc} 
            \toprule
            {Models}  & PSNR$\uparrow$ & SSIM$\uparrow$ \\\midrule
            $B_1$ (w/o RAW in and RAW branch) & 25.79 & 0.915 \\
            $B_2$ (w/o sRGB in and sRGB branch) & 27.24 & 0.929 \\
            $B_3$ (w/o sRGB in) & 26.51 & 0.923 \\
            $B_4$ (w/o RGISP) & \underline{27.38} & \underline{0.932} \\
            $B_5$ (w/o Reconstruction) & 26.93 & 0.926 \\
            $B_6$ (full version) & \textbf{27.88} & \textbf{0.938} \\
            \bottomrule
        \end{tabular}}
\end{table}

Our proposed RRID leverages paired RAW-sRGB data as input. Removing the RAW input and its corresponding RAW demoiréing module ({$B_1$}) results in a drastic decrease in PSNR of approximately 2dB. Likewise, eliminating the sRGB input and its corresponding RGB demoiréing module ($B_2$) leads to a PSNR of only 27.24dB. 
\begin{figure}[!ht]
\centering
\includegraphics[width=0.8\linewidth]{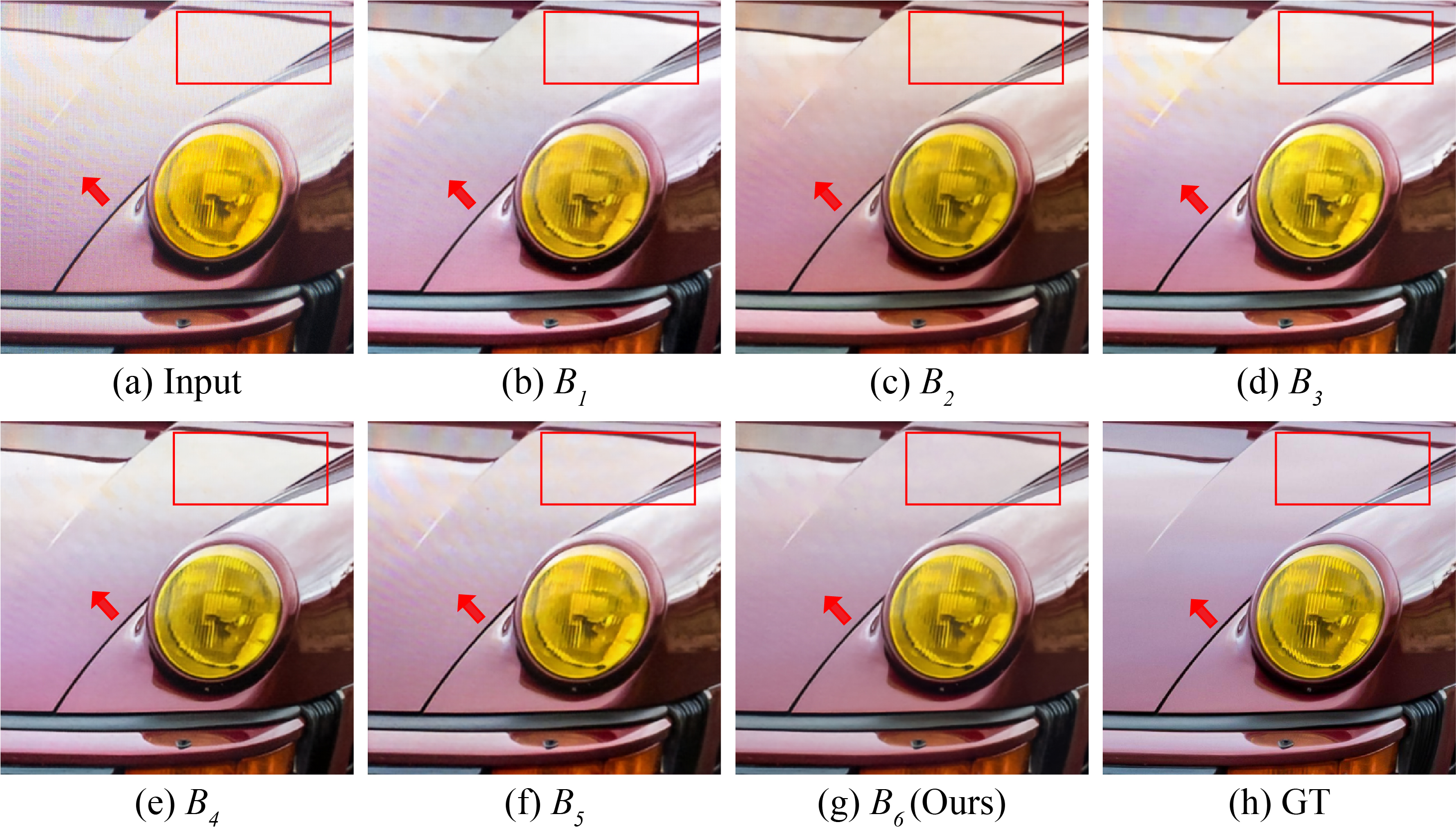}
\caption{Visualization of the ablation study for architecture and inputs comparison.}
\label{fig:ablation_baseline}
\end{figure}
In addition to the decline in numerical metrics, we can also assess the visual performance from Fig.~\ref{fig:ablation_baseline}. When relying only on sRGB data and its corresponding demoiréing modules, moiré patterns are less effectively removed, as depicted in Fig.\ref{fig:ablation_baseline}(b). Conversely, Fig.\ref{fig:ablation_baseline}(c) demonstrates that using only RAW data as input generates images with more pronounced color deviations. To further validate the significance of RAW input for image demoiréing, we compare $B_3$ to $B_1$ by solely removing the RAW input data while retaining the remaining model structures. Although $B_3$ demonstrates some improvements in numerical metrics compared to $B_1$, its performance remains unsatisfactory due to the lack of information provided by the RAW input and the fact that the original RAW demoiréing module is specifically designed for RAW data. Additionally, we conducted further comparisons by removing the originally designed RGISP ($B_4$) and Reconstruction modules ($B_5$), resulting in inferior numerical metrics and performance in Fig.~\ref{fig:ablation_baseline} compared to the complete RRID model ($B_6$).

\noindent\textbf{2) Ablation study on SCDM.}
SCDM is designed to perform effective demoiréing on both RAW and sRGB features. More specifically, we introduce GFM for RAW features demoiréing and FSM for sRGB features demoiréing.

\begin{figure}[!ht]
\centering
\includegraphics[width=1\linewidth]{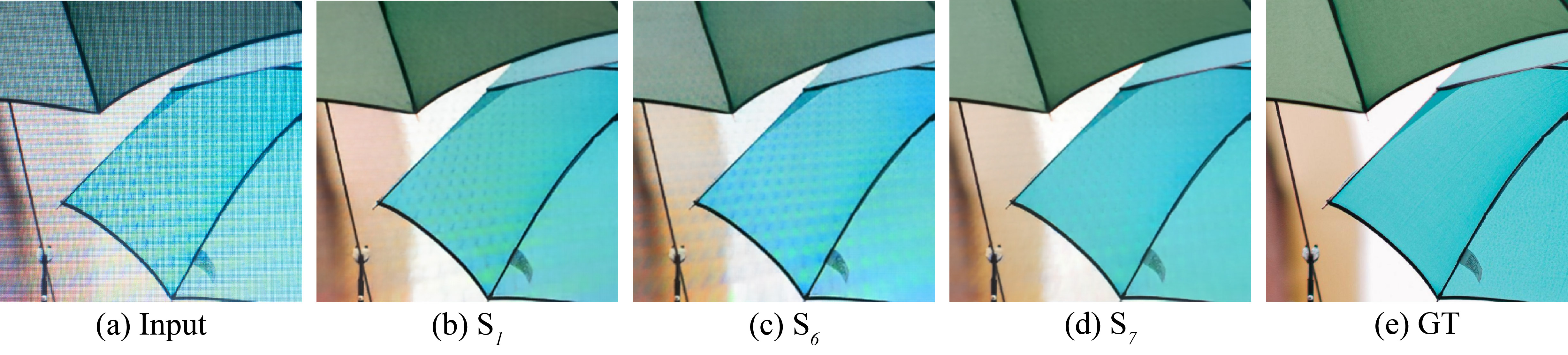}
\caption{Visualization of the ablation study for SCDM.}
\label{fig:ablation_scdm}
\end{figure}

\begin{table}[htbp]
\caption{Ablation study on SCDM.}
\label{table:scdm}
\centering
\scalebox{0.80}{
        \begin{tabular}{lccc} 
            \toprule
            {Models}  & PSNR$\uparrow$ & SSIM$\uparrow$ \\\midrule
            $S_1$ (w/o GFM \& FSM) & 27.00 & 0.926 \\
            $S_2$ (w/o GFM) & 27.17 & 0.927 \\
            $S_3$ (w/o FSM) & 27.39 & 0.930 \\
            $S_4$ (all GFM) & \underline{27.50} & \underline{0.933} \\
            $S_5$ (all FSM) & 27.32 & 0.930 \\
            $S_6$ (w/o DCAB) & 26.55 & 0.923 \\ 
            $S_7$ (SCDM) & \textbf{27.88} & \textbf{0.938} \\
            \bottomrule
        \end{tabular}}
\end{table}
Table~\ref{table:scdm} demonstrates the effectiveness and indispensability of the meticulously designed GFM and FSM integrated within the skip-connections. Removing the demoiréing modules on the skip-connections ($S_1$-$S_3$) individually result in varying degrees of performance decline of the model. For instance, removing both GFM and FSM leads to only 27.00 dB in PSNR.
Furthermore, Fig.\ref{fig:ablation_scdm}(b) demonstrates that the model's ability to remove moiré patterns decreases upon removing the GFM and FSM modules. Furthermore, to assess the specificity of GFM and FSM, we substitute the original modules in $S_4$ and $S_5$, respectively. Despite no reduction in the number of parameters, the PSNR decreases by approximately 0.3-0.5dB. Since DCAB is a crucial component module in SCDM, we substitute DCAB with two convolutional layers to confirm its necessity. Fig.~\ref{fig:ablation_scdm}(c) showcases more residual moiré patterns and significant color deviation compared to our complete SCDM presented in Fig.\ref{fig:ablation_scdm}(d).
\begin{figure}[!ht]
\centering
\includegraphics[width=1\linewidth]{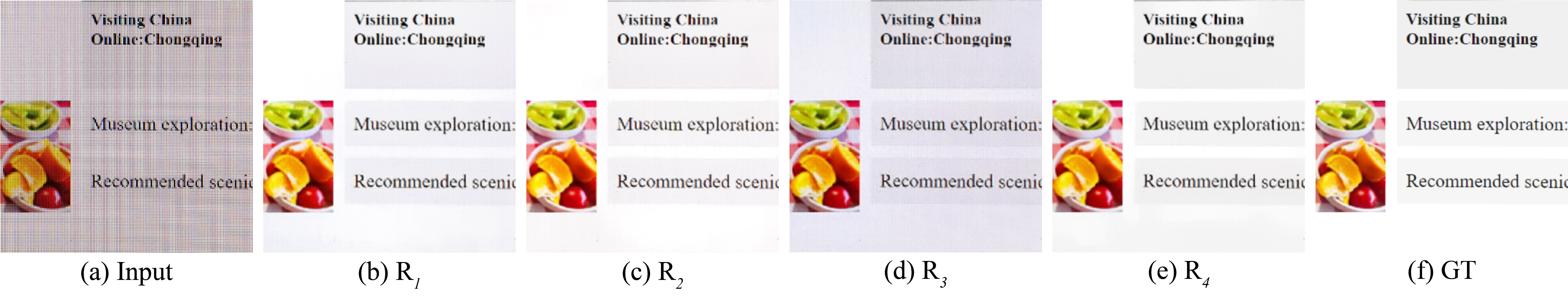}
\caption{Visualization of the ablation study for RGISP.}
\label{fig:ablation_rgisp}
\end{figure}

\noindent\textbf{3) Ablation study on RGISP.}
RGISP is designed to integrate color information from the coarsely demoiréd sRGB features during the ISP stage, facilitating the color recovery process. The process of the RGISP can be recognized as a cross-attention mechanism.
To assess the efficacy of the RGISP design, we compare the following structures in Table~\ref{table:rgisp}. For $R_1$, we concatenate the pre-demoiréd RAW features with the sRGB features and subsequently apply a self-attention mechanism. In $R_2$, the pre-demoiréd RAW features undergo a self-attention mechanism and are then concatenated with the pre-demoiréd sRGB features. Table~\ref{table:rgisp} shows that our proposed RGISP surpasses the self-attention mechanisms ($R_1$, $R_2$), as well as other RAW-to-sRGB conversion methods RRM~\cite{song2023real} ($R_3$).
\begin{table}[htbp]
\caption{Ablation study on RGISP.}
\label{table:rgisp}
\centering
    \scalebox{0.80}{    
        \begin{tabular}{lccc} 
            \toprule
            {Models}  & PSNR$\uparrow$ & SSIM$\uparrow$ \\\midrule
            $R_1$ (self-atten([raw, rgb])) & \underline{27.54} & \underline{0.933} \\
            $R_2$ ([self-atten(raw), rgb]) &  27.42 & 0.931  \\
            $R_3$ (RRM) & 27.21 & 0.929 \\
            $R_4$ (RGISP) & \textbf{27.88} & \textbf{0.938} \\
            \bottomrule
        \end{tabular}}   
\end{table}
In addition, Fig.~\ref{fig:ablation_rgisp} depicts a webpage with globally distributed moiré patterns. $R_1$-$R_4$ are all capable of effectively removing the moiré patterns, but $R_4$ exhibits the best performance in correcting color deviation. With the introduction of sRGB pre-demoiréd features, RGISP is able to learn a device-dependent ISP and restore the original colors of the image effectively.

\section{Conclusion}
\label{sec:con}
In this paper, we propose to jointly utilize RAW and sRGB data for image demoiréing (RRID). 
Firstly, we introduce SCDM with GFM and FSM embedded in skip-connections to handle the demoiréing of RAW and sRGB features, respectively. 
Additionally, we present RGISP to learn a device-dependent ISP, aiding the color recovery process.
Extensive experiments demonstrate the superior performance of our RRID method in both moiré pattern removal and color cast correction. Our approach outperforms state-of-the-art methods by 0.62dB in PSNR and 0.003 in SSIM.

\section*{Acknowledgements}
This work was supported in part by Macau Science and Technology Development Fund under SKLIOTSC-2021-2023, 0072/2020/AMJ, 0022/2022/A, and 0014/2022/AFJ; in part by Research Committee at University of Macau under MYRG-GRG2023-00058-FST-UMDF and  MYRG2022-00152-FST; in part by Natural Science Foundation of Guangdong Province of China under EF2023-00116-FST.

%
%
\bibliographystyle{splncs04}
\bibliography{main}

\newpage
\begin{center}
    \Large \bfseries Supplementary Material:\\Image Demoireing in RAW and sRGB Domains
\end{center}

In this supplementary material, we first conduct the additional ablation study for SCDM in Sec.~\ref{sec:additional_ablation}. Then, a limitation of our proposed RRID is discussed in Sec.~\ref{sec:limitation}. Lastly, more visual comparisons between RRID and several state-of-the-art models are provided in Sec.~\ref{sec:additional_TMM22}.

\setcounter{section}{0}

\section{Additional Ablation Study for SCDM}
\label{sec:additional_ablation}
Table~\ref{table:ablation_SCDM_suppl} presents the effectiveness and efficiency of the special design in SCDM, where the demoiréing modules, GFM and FSM, are embedded in the skip-connections of SCDM.
In this ablation study, we compare the proposed SCDM with the following cases:
\begin{itemize}
\item[$\bullet$] $A_1$. In the RAW branch, DCAB is replaced with GFM, while in the RGB branch, DCAB is replaced with FSM. The demoiréing modules embedded in the skip-connections remain unchanged for both branches.
\item[$\bullet$] $A_2$. In the RAW branch, DCAB is replaced with GFM, while in the RGB branch, DCAB is replaced with FSM. The demoiréing modules embedded in skip-connections are removed for both branches.
\item[$\bullet$] $A_3$. RRID with the original SCDM design.
\end{itemize}

\begin{table}[htbp]
\caption{Ablation Study for SCDM.}
\label{table:ablation_SCDM_suppl}
\centering
        \begin{tabular}{lccc} 
            \toprule
            {Models}  & PSNR$\uparrow$ & SSIM$\uparrow$ & Inference time(s)\\\midrule
            $A_1$ & 27.11 & 0.927 &  4.589 \\
            $A_2$ & 26.68 & 0.923 &  3.909 \\
            $A_3$ (RRID) & \textbf{27.88} & \textbf{0.938} & \textbf{0.089}\\
            \bottomrule
        \end{tabular}
\end{table}

From $A_1$ in Table~\ref{table:ablation_SCDM_suppl}, it can be observed that replacing DCAB with GFM and FSM results in a noticeable decline by 0.77dB in PSNR and 0.011 in SSIM. A more significant concern arises from the substantial increase in inference time to 4.589 seconds, primarily caused by the Block DCT utilized in FSM. This observation underscores the crucial role of DCAB in encoding and decoding informative features as a pivotal block in SCDM. Moreover, even when we remove the demoiréing modules from the skip-connections in $A_1$, the inference time remains relatively long for $A_2$. This observation emphasizes the effectiveness and efficiency of incorporating demoiréing modules within skip-connections.


\section{Limitation}
\label{sec:limitation}
While we propose to utilize paired RAW and sRGB images for moiré pattern removal and design RGISP to learn a device-dependent ISP, performing color restoration in scenes with significant color deviations remains a challenging task.

\begin{figure}[!ht]
\centering
\includegraphics[width=0.8\linewidth]{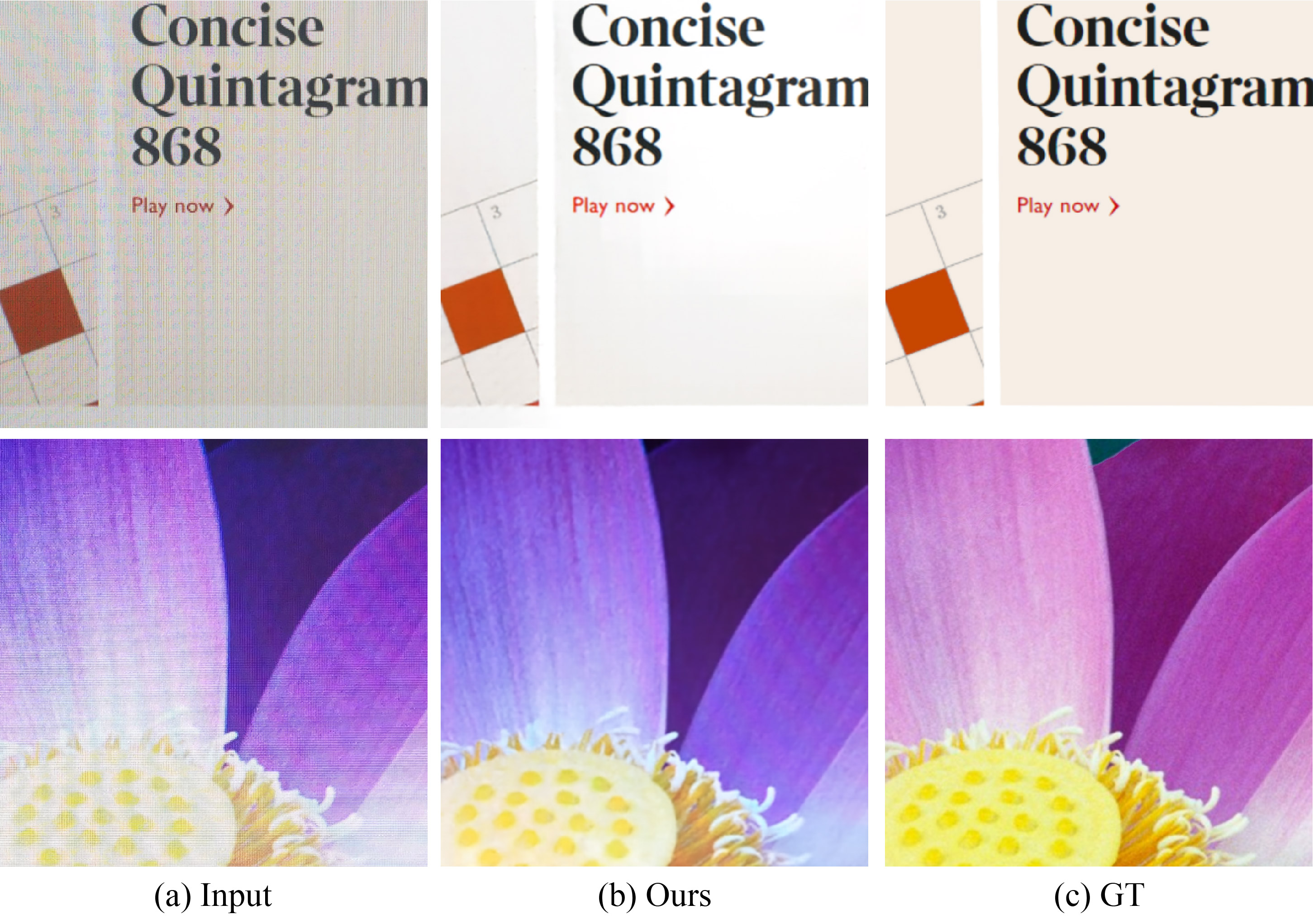}
\caption{Failure cases in correcting color deviations.}
\label{fig:limitation}
\vspace{-5mm}
\end{figure}
Fig.~\ref{fig:limitation} demonstrates two scenes where the moiréd input images exhibit noticeable color discrepancies compared to the ground-truth images. While our proposed method effectively removes moiré patterns and partially corrects the color, it falls short of fully mapping the colors to those present in the ground-truth images.
Currently, there is only one dataset, TMM22 dataset~\cite{yue2022recaptured}, available for image demoiréing with the RAW-sRGB image pairs. However, the images in the training set of TMM22 are limited to a size of 256$\times$256, which may constrain the model's capacity to learn global color mapping. To overcome this limitation, one possible solution is to collect a new paired RAW-sRGB image demoiréing dataset with a higher resolution, which will be our future work.

\section{More Qualitative Comparison on TMM22 Dataset}
\label{sec:additional_TMM22}
\begin{figure*}[!ht]
\centering
\includegraphics[width=0.9\linewidth]{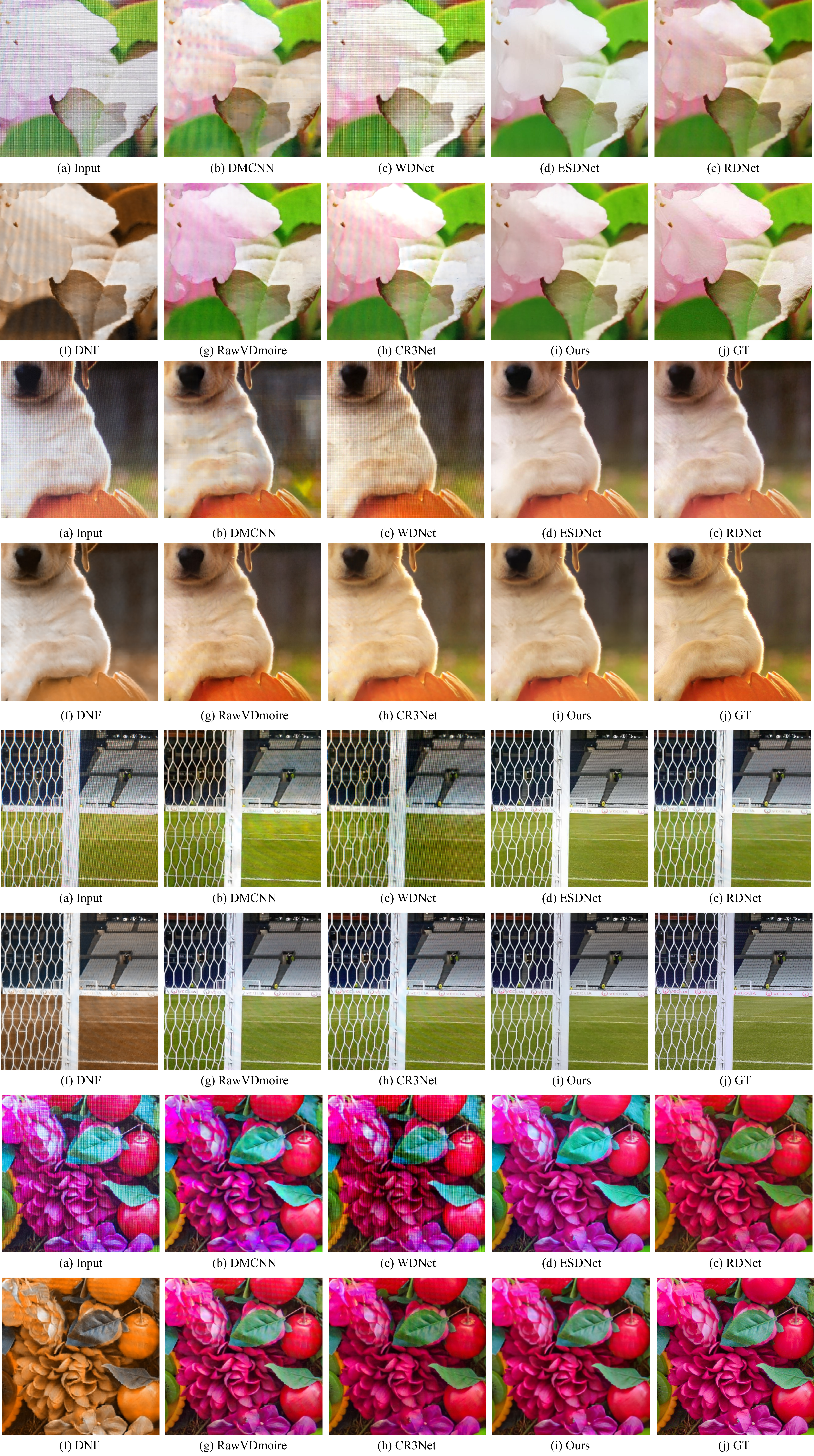}
\caption{Qualitative comparison on raw image demoiréing TMM22 dataset~\cite{yue2022recaptured}.}
\label{fig:TMM22_suppl_1}
\end{figure*}

\begin{figure*}[!ht]
\centering
\includegraphics[width=0.9\linewidth]{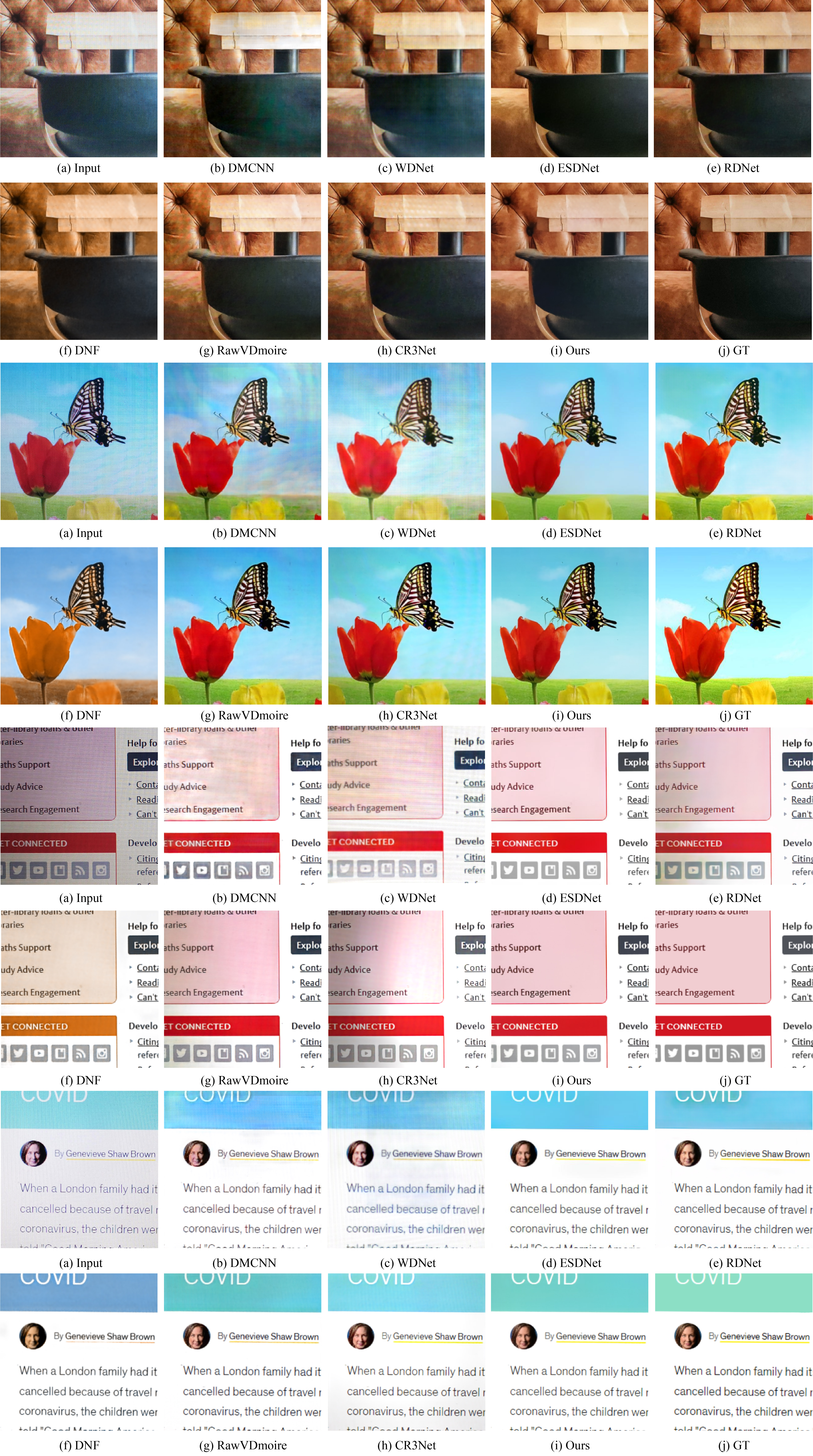}
\caption{Qualitative comparison on raw image demoiréing TMM22 dataset~\cite{yue2022recaptured}.}
\label{fig:TMM22_suppl_2}
\end{figure*}

More visual comparisons with state-of-the-arts methods are presented in Figs.~\ref{fig:TMM22_suppl_1}-\ref{fig:TMM22_suppl_2}. 
The results demonstrate the superiority of our proposed RRID in demoiréing and restoring color. 
For instance, in the first scene depicted in Fig.~\ref{fig:TMM22_suppl_1}, the input image suffers from both severe moiré pattern contamination and noticeable color deviation. Previous methods struggle to remove moiré patterns and encounter difficulties in accurately restoring the petal color to pink. In contrast, our approach successfully eliminates moiré patterns and restores the original color.

The last scene depicted in Fig.~\ref{fig:TMM22_suppl_2} presents a moiréd document. Despite the relatively subtle moiré contamination in this scene, the majority of methods successfully eliminate the moiré patterns. However, the colored area at the top of the image deserves attention, as most methods incorrectly correct it as blue. In contrast, the ground truth image reveals that this area is actually cyan. Remarkably, our method excels in accurately restoring the original color.

\end{document}